\documentclass[epj]{svjour}
\usepackage{graphicx}
\usepackage{amssymb}

\begin{document}

\title{Continuum of many-particle states near the metal-insulator transition in the Hubbard model}

\author{Alexei Sherman}

\institute{Institute of Physics, University of Tartu, W.~Ostwaldi Str 1, 50411 Tartu, Estonia\\
\email{alekseis@ut.ee}}

\date{Received: date / Revised version: date}

\abstract{
The strong coupling diagram technique is used for investigating states near the metal-insulator transition in the half-filled two-dimensional repulsive Hubbard model. The nonlocal third-order term is included in the irreducible part along with local terms of lower orders. Derived equations for the electron Green's function are solved by iteration for moderate Hubbard repulsions and temperatures. Starting iteration from Green's functions of the Hubbard-I approximation with various distances of poles from the real frequency axis continua of different metallic and insulating solutions are obtained. The insulating solutions vary in the width of the Mott gap, while the metallic solutions differ in the shape of the spectral function in the vicinity of the Fermi level. Besides, different scenarios of the metal-insulator transition -- with a sudden onset of a band of mobile states near the Fermi level and with gradual closure of the Mott gap -- are observed with a change in temperature. In spite of these dissimilarities, all solutions have a common curve separating metallic and insulating states in the phase diagram. Near this curve metallic and insulating solutions coexist. For moderate Hubbard repulsions metallic solutions are not Fermi liquids.
\PACS{{xx.xx.xx}{xx}}
}

\maketitle

\section{Introduction}
\label{Intro}
The metal-insulator transition (MIT) in the one-band fer\-mi\-o\-nic Hubbard model was investigated by several methods \cite{Imada}. In the pioneering works of J.~Hubbard \cite{Hubbard63,Hubbard64a,Hubbard64b} equations of motion and uncoupling of Green's functions were used. In \cite{Zaitsev76,Zaitsev78,Goryachev,Izyumov} the MIT was considered by applying the diagram technique for Hubbard operators. Later works employed the dynamic mean-field theory (DMFT) \cite{Georges}, its cluster generalizations \cite{Park,Gull} and the variation cluster approximation \cite{Balzer}. In accord with these works, at temperatures lower than the critical temperature the MIT is the first-order transition with regions of the coexistence of metallic and insulating solutions in the phase diagram.

In this work the strong coupling diagram technique \cite{Vladimir,Vakaru,Metzner,Craco95,Craco96,Moskalenko,Pairault,Craco01,Sherman06a,Sherman06b} is applied for investigating many-particle states in the vicinity of the MIT in the two-dimensional (2D) one-band repulsive Hubbard model. The approach is based on the series expansion of the electron Green's function in powers of the kinetic term of the Hamiltonian. The method has a number of merits. In particular, it is an analytic approach, which permits to carry out calculations in real frequencies, without the need for analytic continuation from the imaginary frequency axis. The obtained equations are solved by iteration, and it is reasonable to use the solution of the lowest order -- the Hubbard-I approximation \cite{Hubbard63,Vladimir} -- as the starting function in the iteration. This function has a free parameter -- the distance $\eta$ of its poles from the real frequency axis, which is introduced to provide proper analytic properties for the retarded Green's function and allows one to check the sensitivity of solutions to a variation of the starting function. There are also a number of disadvantages of the approach, the most important of which is the fact that power expansions of this type are inapplicable for too low temperatures $T$ \cite{Oitmaa}. As follows from calculations, in our case we cannot go below $T\approx 0.2t$, where $t$ is the hopping constant between neighboring sites.

As in a diagram technique with an expansion in powers of a potential energy, in this approach it is possible to separate irreducible diagrams. The sum of all these diagrams is called the irreducible part $K$. It plays the role corresponding to a self-energy in the usual diagram technique. A few lowest-order terms in $K$ are local, that is they do not depend on momentum. In application to the fermionic Hubbard model it was shown \cite{Vladimir,Vakaru,Sherman15a} that the MIT can be described with the irreducible part constructed from two lowest-order local terms. This result reproduces the earlier outcome obtained \cite{Zaitsev76,Zaitsev78} in the diagram technique for Hubbard operators. In this approximation the MIT is the second-order transition. Besides, the approximation was shown \cite{Sherman15b,Sherman15c} to give spectral functions in reasonable agreement with high-temperature results of Monte Carlo simulations for a number of parameters and electron concentrations.

The first nonlocal term, which introduces the momentum dependence into $K$, appears in the third order. It is the first term in an infinite sequence of terms in $K$ with ladder inserts, which describe the interaction of electrons with spin and charge fluctuation \cite{Sherman07,Sherman08,Sherman16b}. In this work the third-order nonlocal term was included in the irreducible part together with local terms of lower orders. The derived equations for the electron Green's function were solved by iteration for moderate values of the Hubbard repulsion $U$ and temperature, as well as for different values of $\eta$ in the starting Green's function obtained from the Hubbard-I approximation. For the local $K$ constructed from lower-order terms a similar procedure gave solutions, which were independent of $\eta$ \cite{Sherman15b,Sherman15c}. We found that the inclusion of the nonlocal term leads to a dependence of obtained solutions on $\eta$, that manifests itself mainly near the Fermi level -- widths of the Mott gap in insulating solutions and shapes of spectral functions in a vicinity of the Fermi level in metallic solutions are changed as $\eta$ varies in some range. Moreover, scenarios of the MIT are also changed with $\eta$ from sudden appearance of a narrow band in the middle of the Mott gap to its gradual closure with increasing $T$. Hence {\sl there exist continua of somewhat different insulating and metallic solutions corresponding to given values of\/} $U$ {\sl and\/} $T$. In spite of the plurality of the solutions, they have a common curve separating metallic and insulating solutions in the $U$-$T$ phase diagram. In this work grand (Landau) potentials of the solutions were not calculated and, therefore, we cannot term the curve as the MIT line. The curve separates solutions obtained by iteration from the Hubbard-I starting functions. We can take one of these solutions and, varying gradually the temperature in iteration, cross the curve. In this way we could obtain metallic solutions in the insulating region of the diagram and insulating solutions in its metallic part. Thus, near the separation curve not only multiplicity of metallic or insulating solutions can be found for the same $U$ and $T$, but also different types of states coexist. This resembles the situation near the first-order transition line observed in DMFT and cluster methods \cite{Georges,Park,Gull,Balzer}. However, the continuous dependence of solutions on $\eta$ implies that there are no gaps between states with the lowest and higher grand potentials. Therefore, more likely our situation can be characterized as a competition of two continua of stationary many-particle states.

Obtained metallic solutions are not Fermi liquids -- for moderate $U$ they exist at comparatively high temperatures and have a finite spectral intensity on the Fermi level. From low temperatures these solutions are separated by the region of insulating states. Such a positioning of the separating curve resembles that obtained in cluster methods \cite{Park,Gull,Balzer} and is in disagreement with the DMFT result \cite{Georges}.

In the local approximation for $K$ it was found \cite{Sherman15b,Sherman15c,Sherman16a} that spectral intensities of both metallic and insulating solutions are suppressed near frequencies $\omega=-\mu$ and $U-\mu$, where for the considered case of half-filling the chemical potential $\mu=U/2$. Analysing the derived equations we came to a conclusion that these pseudogaps arise due to multiple reabsorption of carriers with the creation of states with double site occupancies. Similar to optics, this reabsorption leads to a redistribution in spectral intensities. With the inclusion of the third-order term these pseudogaps are retained in the considered range of parameters. The pseudogaps lead to the distinctive four-band structure of spectral functions, which was noticed earlier in Monte Carlo simulations \cite{Groeber}.

The structure of the paper is the following: in the next section, for convenience, a brief description of the strong coupling diagram technique is given; in Section~\ref{Nonlocal} formulas for the third-order nonlocal term are derived; results of calculations and the main discussion are presented in Section~\ref{SF}; concluding remarks are given in Section~\ref{Concl}.

\section{Strong coupling diagram technique}
\label{SCDT}
For convenience, in this section some notions and results of the strong coupling diagram technique are given. A more detailed discussion can be found in references \cite{Vladimir,Vakaru,Metzner,Craco95,Craco96,Moskalenko,Pairault,Craco01,Sherman06a,Sherman06b}.

The Hamiltonian of the 2D fermionic Hubbard model \cite{Hubbard63} reads
\begin{equation}\label{Hamiltonian}
H=\sum_{\bf ll'\sigma}t_{\bf ll'}a^\dagger_{\bf l'\sigma}a_{\bf l\sigma}
+\frac{U}{2}\sum_{\bf l\sigma}n_{\bf l\sigma}n_{\bf l,-\sigma},
\end{equation}
where 2D vectors ${\bf l}$ and ${\bf l'}$ label sites of a square plane lattice, $\sigma=\pm 1$ is the spin projection, $a^\dagger_{\bf l\sigma}$ and $a_{\bf l\sigma}$ are electron creation and annihilation operators, $t_{\bf ll'}$ is the hopping constant and $n_{\bf l\sigma}=a^\dagger_{\bf l\sigma}a_{\bf l\sigma}$. In this work only the hopping constant $t$ between nearest neighbor sites is supposed to be nonzero.

We shall consider the electron Green's function
\begin{equation}\label{Green}
G({\bf l'\tau',l\tau})=\langle{\cal T}\bar{a}_{\bf l'\sigma}(\tau')
a_{\bf l\sigma}(\tau)\rangle,
\end{equation}
where time dependencies $\bar{a}_{\bf l\sigma}(\tau)=\exp({\cal H}\tau)a^\dagger_{\bf l\sigma}\exp(-{\cal H}\tau)$ and the statistical averaging denoted by the angular bra\-ck\-ets are determined by the operator ${\cal H}=H-\mu\sum_{\bf l\sigma}n_{\bf l\sigma}$,  ${\cal T}$ is the time-or\-de\-r\-ing operator which arranges operators from right to left in ascending order of times $\tau$. In the case of moderate and strong electron correlations, $U\gg t$, for calculating this function the series expansion in powers of the first, kinetic term $H_{\rm k}$ of Hamiltonian (\ref{Hamiltonian}) can be used \cite{Abrikosov},
\begin{eqnarray}\label{series}
&&G({\bf l'\tau',l\tau})=\sum_{k=0}^\infty\frac{(-1)^k}{k!}\int_0^\beta \ldots \int_0^\beta d\tau_1\ldots d\tau_k\nonumber\\
&&\quad\quad\quad\quad\quad\times\langle{\cal T}\bar{a}_{\bf l'\sigma}(\tau') a_{\bf l\sigma}(\tau)H_{\rm k}(\tau_1)\ldots H_{\rm k}(\tau_k)\rangle_{0c},
\end{eqnarray}
where $\beta=1/T$, the subscript 0 near the angular brackets indicates that the averaging and time dependencies of operators are determined by the operator ${\cal H}_0={\cal H}-H_{\rm k}$. The subscript $c$ near the brackets points to the linked-cluster theorem -- terms of the series (\ref{series}), which contain disconnected products of averages, are discarded together with the denominator $\langle\exp(\beta{\cal H}_0)\exp(-\beta{\cal H})\rangle_0$ of the Green's function.

As the Coulomb repulsion in ${\cal H}_0$ is a quartic function of electron operators, Wick's theorem cannot be used for disentangling averages in the power expansion. However, ${\cal H}_0$ is the sum of local terms,
\begin{equation}\label{local}
{\cal H}_0=\sum_{\bf l}{\cal H}_{\bf l},\quad {\cal H}_{\bf l}=\sum_\sigma\bigg(\frac{U}{2}n_{\bf l\sigma}n_{\bf l,-\sigma}-\mu n_{\bf l\sigma}\bigg),
\end{equation}
which allows one to represent any of these averages as a sum of all possible products of local averages of electron operators and respective Kronecker deltas over site indices. The management of these deltas is simplified if the local averages are expressed through local cumulants \cite{Kubo} of electron operators. The simplest example of such a transformation is provided by the average of four operators,
\begin{eqnarray}\label{4operators}
&&\big\langle{\cal T}\bar{a}_{{\bf l}_1\sigma_1}(\tau_1)a_{{\bf l}_2\sigma_2}(\tau_2) \bar{a}_{{\bf l}_3\sigma_3}(\tau_3)a_{{\bf l}_4\sigma_4}(\tau_4)\big\rangle\nonumber\\
&&\quad\quad=C_2(\tau_1\sigma_1,\tau_2\sigma_2,\tau_3\sigma_3,\tau_4\sigma_4) \delta_{{\bf l}_1{\bf l}_2}\delta_{{\bf l}_1{\bf l}_3}\delta_{{\bf l}_1{\bf l}_4}\nonumber\\
&&\quad\quad\quad+C_1(\tau_1\sigma_1,\tau_2\sigma_2) C_1(\tau_3\sigma_3,\tau_4\sigma_4)\delta_{{\bf l}_1{\bf l}_2}\delta_{{\bf l}_3{\bf l}_4}\nonumber\\
&&\quad\quad\quad-C_1(\tau_1\sigma_1,\tau_4\sigma_4) C_1(\tau_3\sigma_3,\tau_2\sigma_2)\delta_{{\bf l}_1{\bf l}_4}\delta_{{\bf l}_2{\bf l}_3},
\end{eqnarray}
where
\begin{eqnarray}\label{cumulants}
&&C_1(\tau_1\sigma_1,\tau_2\sigma_2)=\big\langle{\cal T}\bar{a}_{{\bf l}\sigma_1}(\tau_1)a_{{\bf l}\sigma_2}(\tau_2)\big\rangle,\nonumber\\
&&C_2(\tau_1\sigma_1,\tau_2\sigma_2,\tau_3\sigma_3,\tau_4\sigma_4)\nonumber\\
&&\quad\quad=\big\langle{\cal T}\bar{a}_{{\bf l}\sigma_1}(\tau_1)a_{{\bf l}\sigma_2}(\tau_2) \bar{a}_{{\bf l}\sigma_3}(\tau_3)a_{{\bf l}\sigma_4}(\tau_4)\big\rangle\\
&&\quad\quad\quad-\big\langle{\cal T}\bar{a}_{{\bf l}\sigma_1}(\tau_1)a_{{\bf l}\sigma_2}(\tau_2)\big\rangle\big\langle{\cal T}\bar{a}_{{\bf l}\sigma_3}(\tau_3)a_{{\bf l}\sigma_4}(\tau_4)\big\rangle\nonumber\\
&&\quad\quad\quad+\big\langle{\cal T}\bar{a}_{{\bf l}\sigma_1}(\tau_1)a_{{\bf l}\sigma_4}(\tau_4)\big\rangle\big\langle{\cal T}\bar{a}_{{\bf l}\sigma_3}(\tau_3)a_{{\bf l}\sigma_2}(\tau_2)\big\rangle\nonumber
\end{eqnarray}
are cumulants of the first and second orders. To simplify notations the subscript 0 at the angular brackets is omitted in Eqs.~(\ref{4operators}) and (\ref{cumulants}). All operators in a cumulant belong to the same lattice site, and due to the translation symmetry cumulants of the same order are identical on all lattice sites. The cumulants are calculated for the site Hamiltonian ${\cal H}_{\bf l}$, Eq.~(\ref{local}). In the normal state the numbers of creation and annihilation operators in nonvanishing cumulants and averages have to be equal. As Hamiltonian (\ref{Hamiltonian}) commutes with the total spin projection, sums of these projections in creation and annihilation operators in cumulants have to be equal. Relations between higher-order cumulants and averages can be found in \cite{Kubo}.

Thus, averages in the series (\ref{series}) are presented as sums of all possible products of cumulants of different orders and hopping constants $t_{\bf ll'}$ with site indices corresponding to the cumulants. Signs of terms in these sums are determined by the number of permutations, which are necessary to obtain the order of fermion operators in a concrete term (see Eq.~(\ref{4operators})). As in the usual diagram technique with a quadratic unperturbed Hamiltonian \cite{Abrikosov}, prefactors $1/k!$ in (\ref{series}) are removed together with the exclusion of topologically equivalent terms, which differ only by permutations of $k$ operators $H_{\rm k}(\tau_i)$ in (\ref{series}). This opens the way for the partial summation of series terms, since after exclusion of this prefactor a term of the series becomes a product of multipliers, which are independent of its order. There is, however, one essential difference from the usual diagram technique -- terms, which contain cumulants of the second and higher orders and differ only in permutation of two or more creation or annihilation operators in the same cumulant, have to be considered as one term. As a result terms get fractional prefactors.

Terms of the series (\ref{series}) may be presented graphically -- cumulants are depicted by circles connected by lines corresponding to hopping integrals. The lines are directed, say, from a creation operator in a cumulant to an annihilation operator in the same or another cumulant. Thus, the number of incoming or outgoing lines indicates the cumulant order. Among all diagrams it is convenient to set off irreducible diagrams, which cannot be separated into two disconnected parts by cutting a hopping line. Other diagrams can be constructed from two or more irreducible diagrams and connecting lines. In an expression corresponding to such a combined diagram prefactors of its irreducible components are multiplied. The series (\ref{series}) looks like a sum of chains of all possible lengths, constructed from irreducible diagrams and connecting them hopping lines. It is convenient to introduce a notion of the irreducible part $K$ as a sum of all irreducible diagrams. With this quantity the series (\ref{series}) becomes the sum of the irreducible part, two, three and so on irreducible parts connected by the respective number of hopping lines. After the Fourier transformation to 2D wave vectors ${\bf k}$ and Matsubara frequencies $\omega_j=(2j-1)\pi T$ with an integer $j$ this sum can be rewritten as
\begin{equation}\label{Larkin}
G({\bf k}j)=\Big\{\big[K({\bf k}j)\big]^{-1}-t_{\bf k}\Big\}^{-1}.
\end{equation}

\begin{figure}[t]
\centerline{\resizebox{0.99\columnwidth}{!}{\includegraphics{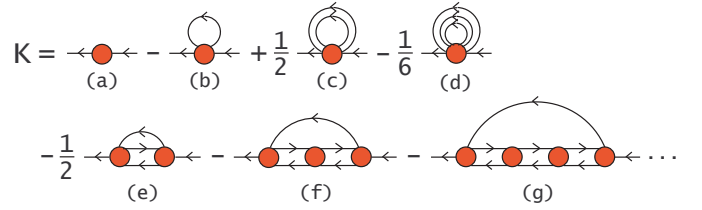}}}
\caption{Diagrams of several lowest orders in $K({\bf k}j)$.} \label{Fig1}
\end{figure}

Diagrams of several lowest orders in $K({\bf k}j)$ are shown in Fig.~\ref{Fig1} with their signs and prefactors.
Thanks to the possibility of partial summation, bare internal lines $t_{\bf k}$ in Fig.~\ref{Fig1} can be transformed into dressed ones,
\begin{equation}\label{hopping}
\theta({\bf k}j)=t_{\bf k}+t^2_{\bf k}G({\bf k}j).
\end{equation}

\section{The lowest-order nonlocal term}
\label{Nonlocal}
As was shown in \cite{Vladimir,Vakaru,Sherman15a,Sherman15b,Sherman15c}, $K({\bf k}j)$ approximated by the sum of diagrams (a) and (b) in Fig.~\ref{Fig1} allows one to describe the MIT and to obtain spectral functions in reasonable agreement with Monte Carlo results for $T\gtrsim t$. In this approximation the irreducible part is local, $K({\bf l'\tau',l\tau})\propto\delta_{\bf l'l}$, and, therefore, its Fourier transform is independent of ${\bf k}$. The diagrams (c) and (d) correspond to local terms and, therefore, they are supposed not to introduce qualitative changes into the solutions obtained earlier. The lowest-order nonlocal term is given by the diagram (e). It is the first term in the sequence of diagrams with ladder inserts, some of which are shown in the second row in Fig.~\ref{Fig1}. The sums of ladder diagrams describe spin and charge susceptibilities \cite{Sherman07,Sherman08,Sherman16b}, and the diagrams with their inserts take into account interactions of electrons with spin and charge fluctuations.

In this work we consider changes in spectral functions introduced by the inclusion of the lowest-order nonlocal term (e) in $K({\bf k}j)$, in addition to the diagrams (a) and (b). In this case the irreducible part reads
\begin{eqnarray}\label{K}
&&K({\bf k}j)=C_1(j)-\frac{T}{N}\sum_{{\bf k'}j'\sigma'} C_2(j\sigma,j\sigma,j'\sigma',j'\sigma')\theta({\bf k'}j')\nonumber\\
&&\quad-\frac{T^2}{2N^2}\sum_{{\bf k'k''}j'j''}\theta({\bf k+k''},j+j'')\theta({\bf k'+k''},j'+j'') \nonumber\\
&&\quad\quad\quad\quad\times\theta({\bf k'}j')\Big[C_2(j\sigma;j+j'',\sigma;j'+j'',-\sigma;j',-\sigma)\nonumber\\
&&\quad\quad\quad\quad\quad\quad\quad\times C_2(j+j'',\sigma;j\sigma;j',-\sigma;j'+j'',-\sigma)\nonumber\\
&&\quad\quad\quad\quad\quad\quad\quad+\sum_{\sigma'} C_2(j\sigma;j+j'',\sigma';j'+j'',\sigma';j'\sigma)\nonumber\\
&&\quad\quad\quad\quad\quad\quad\quad\times C_2(j+j'',\sigma';j\sigma;j'\sigma;j'+j'',\sigma')\Big],
\end{eqnarray}
where the abbreviation $j+j'$ refers to the frequency $\omega_{j+j'}$ $=[2(j+j')-1]\pi T$, four frequencies and spin projections in $C_2$, of which only three frequencies and projections are independent, belong to four legs of the cumulant, $N$ is the number of sites.

Equations describing second-order cumulants are given in \cite{Moskalenko,Sherman06a,Sherman06b,Sherman07,Sherman08}. They are rather cumbersome. However, the equations can be significantly simplified in the case
\begin{equation}\label{condition}
T\ll\mu,\quad T\ll U-\mu.
\end{equation}
For $U\gg T$ this range of $\mu$ contains most interesting cases of half-filling, $\mu=U/2$, and moderate doping. For the conditions (\ref{condition}) the first- and second-order cumulants in (\ref{K}) read
\begin{eqnarray}\label{C12}
&&C_1(j)=\frac{1}{2}[g_1(j)+g_2(j)],\nonumber \\
&&C_2(j\sigma;j+j'',\sigma;j'+j'',-\sigma;j',-\sigma)=-\bigg(\frac{\delta_{jj'}}{2T} +\frac{\delta_{j''0}}{4T}\bigg)\nonumber \\
&&\quad\quad\times F(j+j'')F(j')+B(jj'j''),\nonumber \\[-0.5ex]
&&\\[-0.5ex]
&&C_2(j\sigma;j+j'',\sigma';j'+j'',\sigma';j',\sigma)\nonumber\\
&&\quad=\frac{1}{4T}\big[\delta_{jj'}\big(1-2\delta_{\sigma\sigma'}\big)+
\delta_{j''0}\big(2-\delta_{\sigma\sigma'}\big)\big]\nonumber\\
&&\quad\quad\times F(j+j'')F(j')- \delta_{\sigma,-\sigma'}B(jj'j''), \nonumber
\end{eqnarray}
where
\begin{eqnarray}\label{terms}
&&g_1(j)=(i\omega_j+\mu)^{-1},\quad g_2=(i\omega_j+\mu-U)^{-1},\nonumber\\
&&F(j)= g_1(j)-g_2(j),\nonumber \\
&&B(jj'j'')=\frac{1}{2}F(j')\big[g_1(j+j'')g_1(j'+j'')\nonumber\\[-0.5ex]
&&\\[-0.5ex]
&&\quad\quad+g_2(j+j'')g_2(j)-g_1(j'+j'')g_2(j)\big] \nonumber\\
&&\quad+\frac{1}{2}F(j+j'')\big[g_1(j)g_1(j')\nonumber\\
&&\quad\quad+g_2(j')g_2(j'+j'')-g_1(j)g_2(j'+j'')\big].\nonumber
\end{eqnarray}

It is worth noting that $C_2$ in (\ref{C12}) contain terms proportional to $1/T$. The divergence of these terms at $T\rightarrow 0$ indicates that the used approach is not applicable to very low temperatures, which is the known defect of such series expansions \cite{Oitmaa}. Indeed, we found that there appear large regions with negative intensities in spectral functions for $T\lesssim 0.2t$, while for larger temperatures general shapes of spectra are similar to those obtained in Monte Carlo \cite{Groeber} and our earlier calculations \cite{Sherman15b,Sherman15c}. With this observation and taking into account the condition (\ref{condition}) we limited the considered temperature range to
\begin{equation}\label{trange}
0.3t\lesssim T\lesssim U/6.
\end{equation}
for half-filling.

After substituting Eqs.~(\ref{C12}) and (\ref{terms}) in (\ref{K}) the irreducible part reads
\begin{eqnarray}\label{Kfin}
&&K({\bf k}j)=\frac{1}{2}[g_1(j)+g_2(j)]+F^2(j)\frac{3}{4N}\sum_{\bf k'}\theta({\bf k'}j)\nonumber \\
&&\quad-F(j)\frac{T}{2N}\sum_{{\bf k'}j'}\theta({\bf k'}j')L(j')
-L(j)\frac{T}{2N}\sum_{{\bf k'}j'}\theta({\bf k'}j')F(j')\nonumber \\
&&\quad-F^4(j)\frac{3}{16N^2}\sum_{\bf k'k''}\theta({\bf k'}j)\theta({\bf k+k''},j)\theta({\bf k'+k''},j)\nonumber\\
&&\quad+F(j)\bigg[2L(j)-\frac{F(j)}{T}\bigg]\frac{3T}{8N^2}\sum_{{\bf k'k''}j'}\theta({\bf k'}j')\theta({\bf k+k''},j)\nonumber\\
&&\quad\quad\times\theta({\bf k'+k''},j')F^2(j')\nonumber\\
&&\quad+F(j)^2\frac{3T}{4N^2}\sum_{{\bf k'k''}j'}\theta({\bf k'}j')\theta({\bf k+k''},j)
\theta({\bf k'+k''},j')\nonumber\\
&&\quad\quad\times F(j')L(j')\nonumber\\
&&\quad-\frac{T^2}{N^2}\sum_{{\bf k'k''}j'j''}\theta({\bf k'}j')\theta({\bf k+k''},j+j'') \nonumber\\
&&\quad\quad\times\theta({\bf k'+k''},j'+j'')B(jj'j'')B(j'jj''),
\end{eqnarray}
where $L(j)=F(j)g_1(j)+g_2^2(j)$.

In view of the unreliability of methods of the numeric continuation to the real frequency axis, performed in do\-u\-b\-le- or even in quadruple-precision arithmetics \cite{Beach,Schoetta,Schoettb}, calculations were carried out in real frequencies from the very beginning. In Eq.~(\ref{Kfin}) the transition $i\omega_j\rightarrow\omega$ is easily accomplished in all terms except the last one, in which $i\omega_j$ enters in combinations with summation variables and, therefore, cannot be simply substituted with $\omega$. To simplify this term the renormalized hopping $\theta({\bf k}j)$ was replaced by the bare one $t_{\bf k}$. Even with this simplification the double summation over $j'$ and $j''$ cannot be carried out analytically in this term. To overcome this difficulty we took into account that $T$ is the smallest energy parameter in these summations and transformed them into a double integral, which can be calculated,
\begin{eqnarray}\label{lastterm}
&&T^2\sum_{j'j''}B(jj'j'')B(j'jj'')\nonumber\\
&&\quad\approx\frac{1}{(2\pi)^2}\int\!\!\!\! \int_{-\infty}^{\infty} B(i\omega_j,iy,ix)B(iy,i\omega_j,ix)dx\,dy\nonumber\\
&&\quad=\frac{3}{2}g_1^2(j)g_2^2(j).
\end{eqnarray}
This approximation works well for very small $T$. For larger $T$ arising smallish differences between the first and third rows of Eq.~(\ref{lastterm}) were compensated by introducing a factor into the last expression, the value of which was determined by comparing real parts of the first and third rows at the Matsubara frequency $-\pi T$.

Calculations of the electron Green's function from the set of equations (\ref{Larkin}), (\ref{hopping}), (\ref{Kfin}) and (\ref{lastterm}) were performed by iteration. As the starting expression in this procedure we used the Green's function of the Hubbard-I approximation \cite{Hubbard63}, which is obtained with the irreducible part reduced to the first term in the right-hand side of (\ref{Kfin}) \cite{Vladimir}. To provide the initial function with proper analytic properties for a retarded Green's function and to get finite peak widths in its spectral function, the poles in the initial function were shifted from the real frequency axis to the lower half-plane by the distance $\eta$. Values of $\eta$ varied from $0.16t$ to $1.1t$. A further increase of $\eta$ does not change solutions. The lower limit for $\eta$ is conditioned by the requirement that its value be several times larger than the frequency discretization step used in calculations (otherwise the initial function were ill described). The least used frequency discretization step was $0.04t$.

\section{Spectral functions}
\label{SF}
\begin{figure}[t]
\centerline{\resizebox{0.8\columnwidth}{!}{\includegraphics{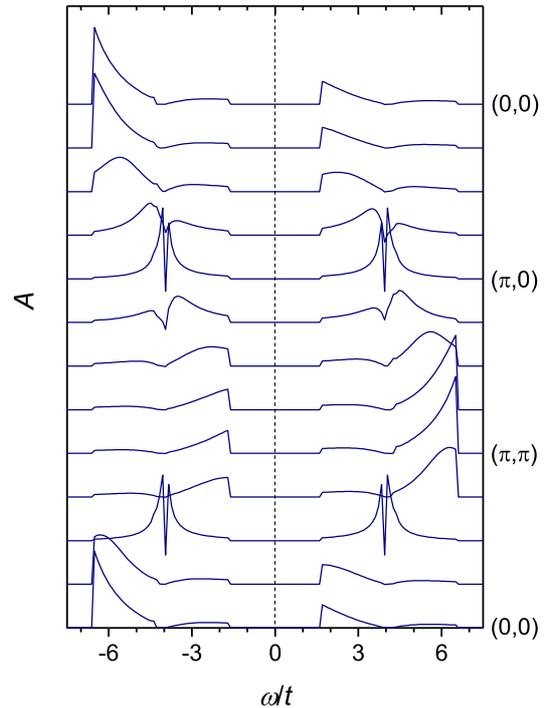}}}
\caption{The spectral functions along the symmetry lines of the Brillouin zone for half-filling, $U=8t$, $T=0.691t$ and $\eta=1.07t$. Momenta corresponding to neighbor functions differ by fixed steps along the lines. Momenta of functions in the symmetric points are indicated near the right axis. The zero frequency corresponds to the Fermi level.} \label{Fig2}
\end{figure}

\begin{figure}[t]
\centerline{\resizebox{0.8\columnwidth}{!}{\includegraphics{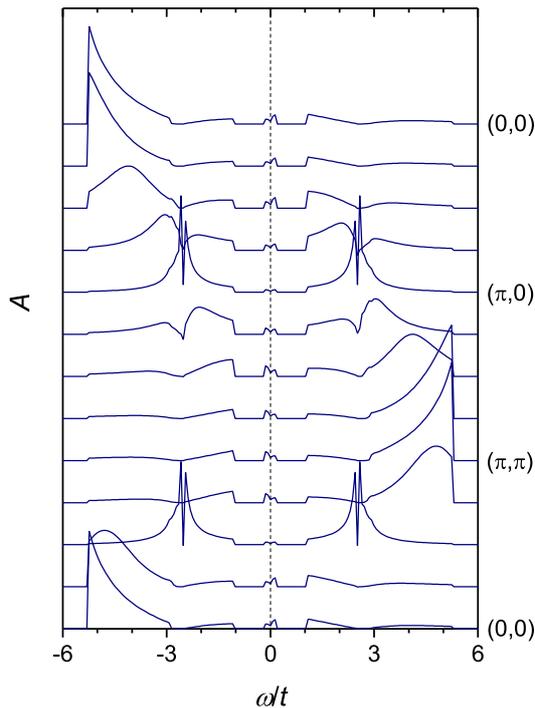}}}
\caption{Same as in Fig.~\protect\ref{Fig2}, but for $U=5.1t$, $T=0.633t$ and $\eta=0.408t$.} \label{Fig3}
\end{figure}

\begin{figure}[t]
\centerline{\resizebox{0.8\columnwidth}{!}{\includegraphics{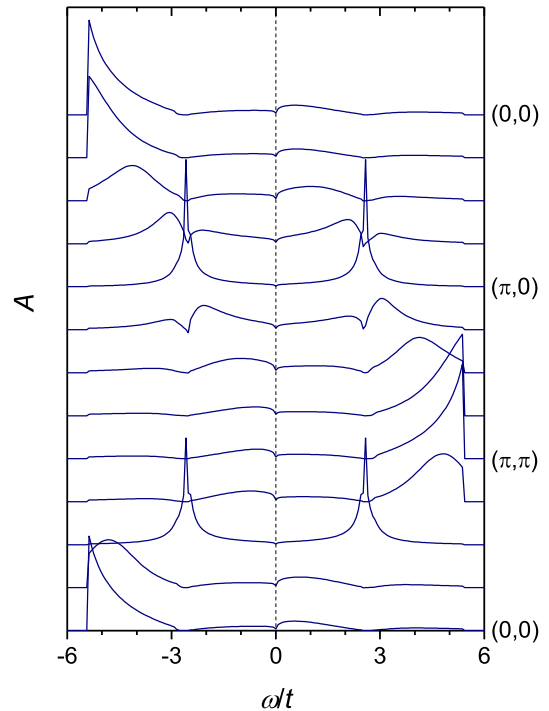}}}
\caption{Same as in Fig.~\protect\ref{Fig2}, but for $U=5.1t$, $T=0.633t$ and $\eta=0.68t$.} \label{Fig4}
\end{figure}

In this section, spectral functions $$A({\bf k}\omega)=-\pi^{-1}{\rm Im}\, G({\bf k}\omega)$$ of solutions of the above equations are given in the case of half-filling, $\mu=U/2$. Three typical dependencies of their shapes on the momentum varying along symmetry lines of the Brillouin zone are shown in Figs.~\ref{Fig2}--\ref{Fig4}. These and following results were obtained in an 8$\times$8 lattice. Some trial calculations were carried out also for a 12$\times$12 lattice. Their results for respective wave vectors appeared to be close to those obtained in the smaller lattice. Below the lattice spacing is taken as the unit of length.

Parameters of Fig.~\ref{Fig2} correspond to an insulating case, with a broad Mott gap around the Fermi level. Comparing these spectra with results obtained for the same parameters in lower-order approximation \cite{Sherman15b,Sherman15c} one can see that the gap in Fig.~\ref{Fig2} is broader, spectra are sharper and the reabsorption pseudogaps at $\omega=\pm U/2$ are more conspicuous. As discussed in \cite{Sherman15b,Sherman15c,Sherman16a}, these pseudogaps arise due to multiple reabsorption of carriers with the creation of states with double occupancy of sites. The pseudogaps lead to the characteristic four-band structure of the spectra, which was first observed in Monte Carlo simulations \cite{Groeber}. As seen in Fig.~\ref{Fig2}, near the pseudogap frequencies small regions of negative intensity may appear. As a whole shapes and locations of maxima in spectra in Fig.~\ref{Fig2} are similar to those obtained in lower-order approximation \cite{Sherman15b,Sherman15c} and by cluster approaches \cite{Senechal,Dahnken,Kyung}.

\begin{figure}[tbh]
\centerline{\resizebox{0.8\columnwidth}{!}{\includegraphics{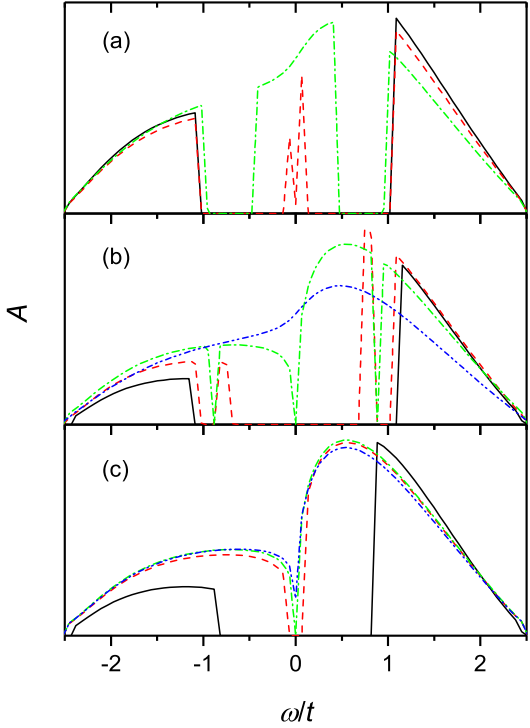}}}
\caption{Three scenarios of the metal-insulator transition. Spectral functions near the Fermi level for half-filling, ${\bf k}=(0,0)$ and $U=5.1t$. (a) $\eta=0.408t$, $T=0.611t$ (black solid line), $T=0.619t$ (red dashed line) and $T=1.101t$ (green dash-dotted line). (b) $\eta=0.544t$, $T=0.33t$ (black solid line), $T=0.44t$ (red dashed line), $T=0.605t$ (green dash-dotted line) and $T=1.651t$ (blue dash-dot-dotted line). (c) $\eta=0.68t$, $T=0.33t$ (black solid line), $T=0.578t$ (red dashed line), $T=0.605t$ (green dash-dotted line) and $T=0.633t$ (blue dash-dot-dotted line).} \label{Fig5}
\end{figure}

Figures~\ref{Fig3} and \ref{Fig4} correspond to two qualitatively different metallic solutions, which have finite spectral intensities on the Fermi level. These solutions differ not only in shapes of spectral functions near the Fermi level, but also in the character of MIT, that is demonstrated in Fig.~\ref{Fig5}. Figure~\ref{Fig5}(a) shows the temperature variation of the ${\bf k}=(0,0)$ spectrum near $\omega=0$ for the same $U$ and $\eta$ as in Fig.~\ref{Fig3}. As the temperature rises, a narrow band appears suddenly close to the Fermi level in the Mott gap, pointing to the transition to a metallic state. Though this behavior resembles to some extend MIT in DMFT \cite{Georges}, the band is not an analogue of the Abrikosov-Suhl peak -- in the used approach all electrons are identical. As seen from Fig.~\ref{Fig3}, the shape of the band depends on the wave vector, that implies mobile character of the states. With further increase in temperature the band broadens out and finally fills the gap completely. Figure~\ref{Fig5}(c), which corresponds to the same $U$ and $\eta$ as in Fig.~\ref{Fig4}, demonstrates another scenario of MIT. Here a metallic state occurs by a gradual closure of the Mott gap as the temperature increases. This behavior resembles that found in lower-order approximation \cite{Sherman15a,Sherman15b,Sherman15c}. Finally, Fig.~\ref{Fig5}(b) shows one more case of the MIT, which has some features similar to those of the previous two cases. Here the formation of a band in the Mott gap starts from its peripheral parts, which are broadened in the direction to the Fermi level with increasing temperature and eventually fill the gap.

\begin{figure}[tbh]
\centerline{\resizebox{0.8\columnwidth}{!}{\includegraphics{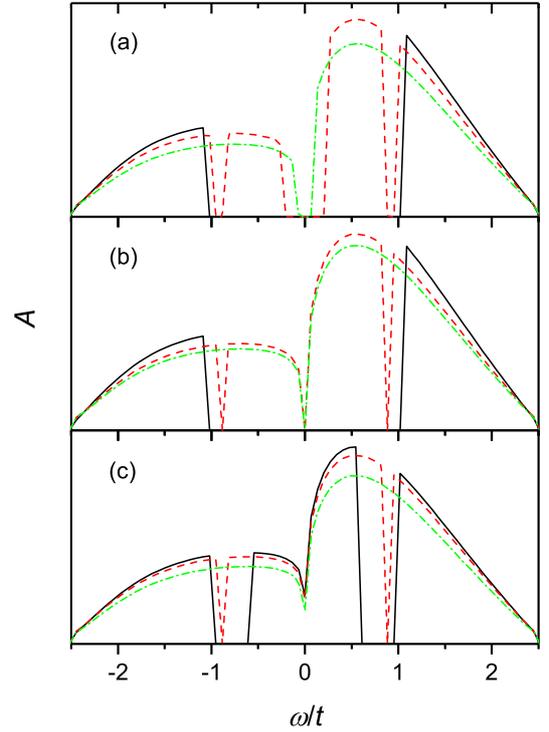}}}
\caption{Dependencies of the spectral functions near the Fermi level on $\eta$ for $U=5.1t$, ${\bf k}=(0,0)$ and different temperatures. (a) $T=0.578t$, insulator states, $\eta=0.476t$ (black solid line), $\eta=0.544t$ (red dashed line) and $\eta=0.68t$ (green dash-dotted line). (b) $T=0.605t$, insulator and critical states, $\eta=0.408t$ (black solid line), $\eta=0.544t$ (red dashed line) and $\eta=0.68t$ (green dash-dotted line). (c) $T=0.633t$, metallic states, $\eta=0.476t$ (black solid line), $\eta=0.544t$ (red dashed line) and $\eta=0.68t$ (green dash-dotted line).}\label{Fig6}
\end{figure}

It should be emphasized that solutions in Figs.~\ref{Fig3} and \ref{Fig4} were obtained for the same values of $U$ and $T$, however for different values of the pole shift $\eta$ in the starting Green's function of the iteration process. That is, {\em Green's and spectral functions depend heavily on the starting function and vary continuously with\/} $\eta$. Main changes in spectra caused by variation of $\eta$ occur near the Fermi level, while the periphery of spectral functions is affected only slightly. Spectra of the type shown in Figs.~\ref{Fig3} and \ref{Fig5}(a) are obtained for smaller values of $\eta$, for $U=5.1t$ at $\eta\approx 0.27t-0.45t$. Spectra similar to those depicted in Fig.~\ref{Fig4} and \ref{Fig5}(c) result from larger $\eta$, for $U=5.1t$ at $\eta\gtrsim 0.68t$. Spectra of the type of Fig.~\ref{Fig5}(b) are observed for intermediate $\eta$. The mentioned variation of spectra with changing $\eta$ is inherent in the intermediate range of Hubbard repulsions $4.5t\lesssim U\lesssim 6t$. Outside of this range variations in spectra are less pronounced. It should be underlined that in the lower-order approximation \cite{Sherman15b,Sherman15c} spectra was independent of $\eta$. Thus, this dependance is caused by the nonlocal term (e) in Fig.~\ref{Fig1}. Examples of the variation of spectral functions with $\eta$ are shown in Fig.~\ref{Fig6}.

Hence, as the shift $\eta$ varies continuously, there exists {\sl a continuum of many-particle solutions} for the set of equations (\ref{Larkin}), (\ref{hopping}), (\ref{Kfin}) and (\ref{lastterm}) for given values of $U$ and $T$. Since Green's function is an image of a many-particle state, one can say about a continuum of such states. Evidently they correspond to different values of the grand potential, and the continuity of the dependence of spectral function shapes on $\eta$ implies the continuity of the respective potentials. In its turn, it suggests that the metallic or insulating solution, which for given $U$ and $T$ has the smallest value of the grand potential, is not separated by a gap from states with larger potentials. In the used approximation all them are stationary states, and a crystal can be brought into any of them in the course of preparation. These states are supposed to transform to long-living metastable states under perturbations. The fact that the solutions are many-particle states of a macroscopic number of electrons, which are connected by strong correlations, counts in favour of this supposition -- perceptible changes in such states need in many-particle perturbations. The continuity of the solutions, caused by the nonlocal term, suggests that the inclusion of unaccounted nonlocal diagrams in $K$ will transform these solutions into a continuum of somewhat different stationary or long-living metastable states. The results were obtained for finite temperatures. Therefore, the above-mentioned states are mixed states constructed from pure states -- eigenstates of Hamiltonian (\ref{Hamiltonian}) -- with different weighting factors.

\begin{figure}[tbh]
\centerline{\resizebox{0.98\columnwidth}{!}{\includegraphics{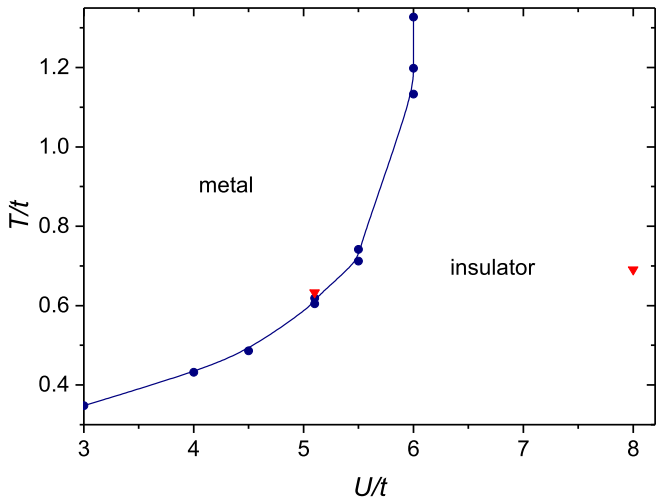}}}
\caption{The phase diagram. The curve separates metallic and insulating states obtained by iteration from Hubbard-I solutions with different $\eta$. Circles indicate parameters, for which critical solutions of the type shown in Fig.~\protect\ref{Fig6}(b) were obtained in calculations. Triangles show parameters of Figs.~\protect\ref{Fig2}--\protect\ref{Fig4}.} \label{Fig7}
\end{figure}

In spite of the distinctions in spectral function shapes obtained with different $\eta$, for all values of this parameter there exist a common curve on the $U$-$T$ phase diagram, which separates insulating and metallic solutions. This curve is shown in Fig.~\ref{Fig7} for the temperature range defined by inequalities (\ref{trange}). It should be noted that the curve separates solutions obtained by iteration from the Hubbard-I Green's function. We can take one of these solutions as the initial one and change gradually the temperature in iteration to cross the curve. It turns out that in such a manner it is possible to obtain metallic solutions in the insulating region and insulating solutions in the metallic region at some distance from the curve in Fig.~\ref{Fig7}. This result resembles to some extent the behavior at the first-order MIT observed in DMFT \cite{Georges} and cluster approaches \cite{Park,Gull,Balzer}. However, in our case it would be better to say about two competing continua of solutions rather than two solutions because, as was mentioned above, metallic and insulating states with lowest grand potentials are not separated from other states. It is worth noting that the separation curve in Fig.~\ref{Fig7} runs from smaller $T$ and $U$ to the region of larger values of these parameters. The same orientation of the first-order transition line was obtained in cluster approaches \cite{Park,Gull,Balzer} and the opposite one -- from larger $T$ and smaller $U$ to smaller $T$ and larger $U$ -- in DMFT \cite{Georges}. It should be noticed that grand potentials of solutions were not calculated in this work and, therefore, the exact location of the transition line is unknown at present. However, it is supposed to be close to the separation line. We did not find a critical point in the range (\ref{trange}).

\begin{figure}[tbh]
\centerline{\resizebox{0.8\columnwidth}{!}{\includegraphics{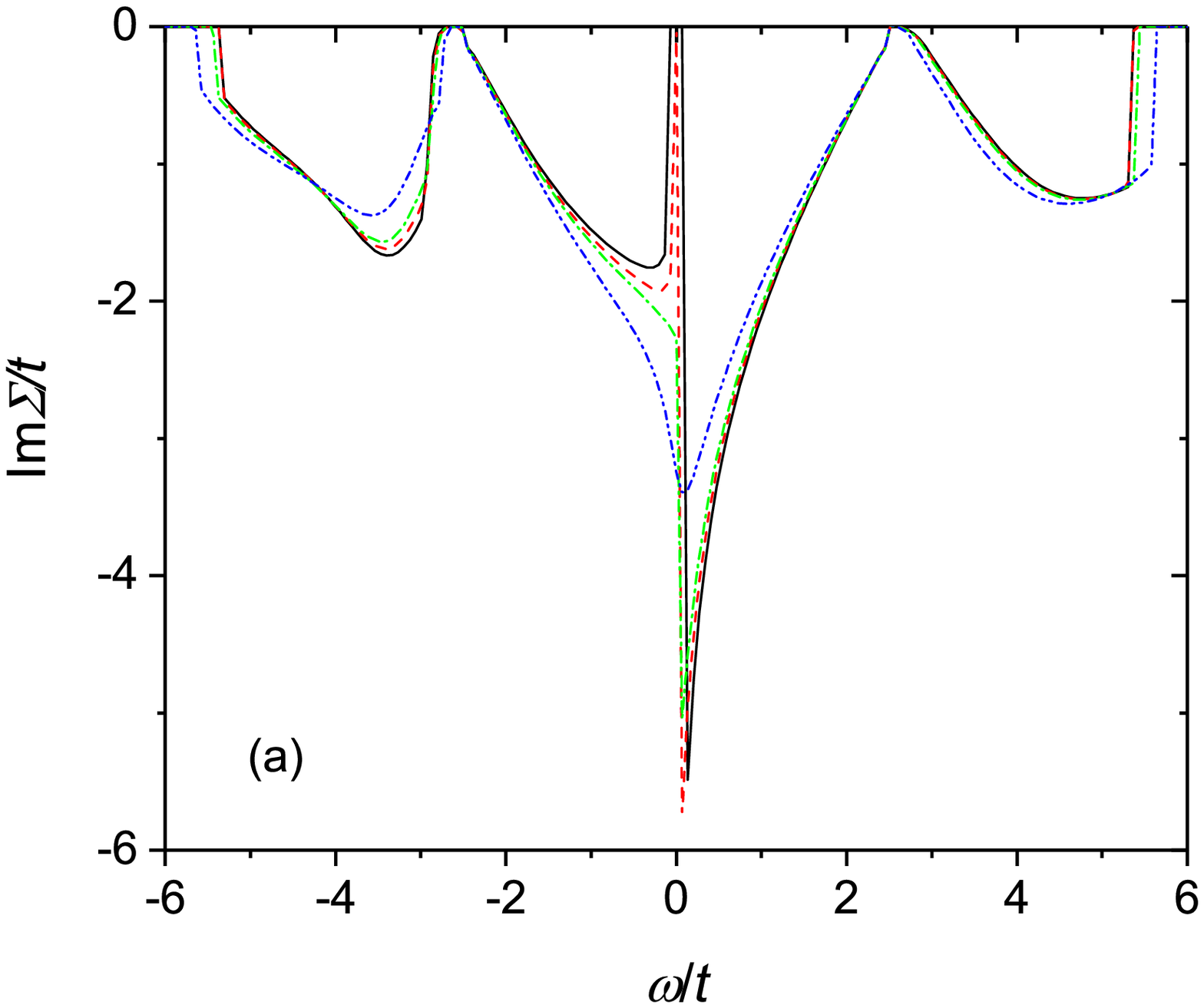}}}
\vspace*{3ex}
\centerline{\resizebox{0.8\columnwidth}{!}{\includegraphics{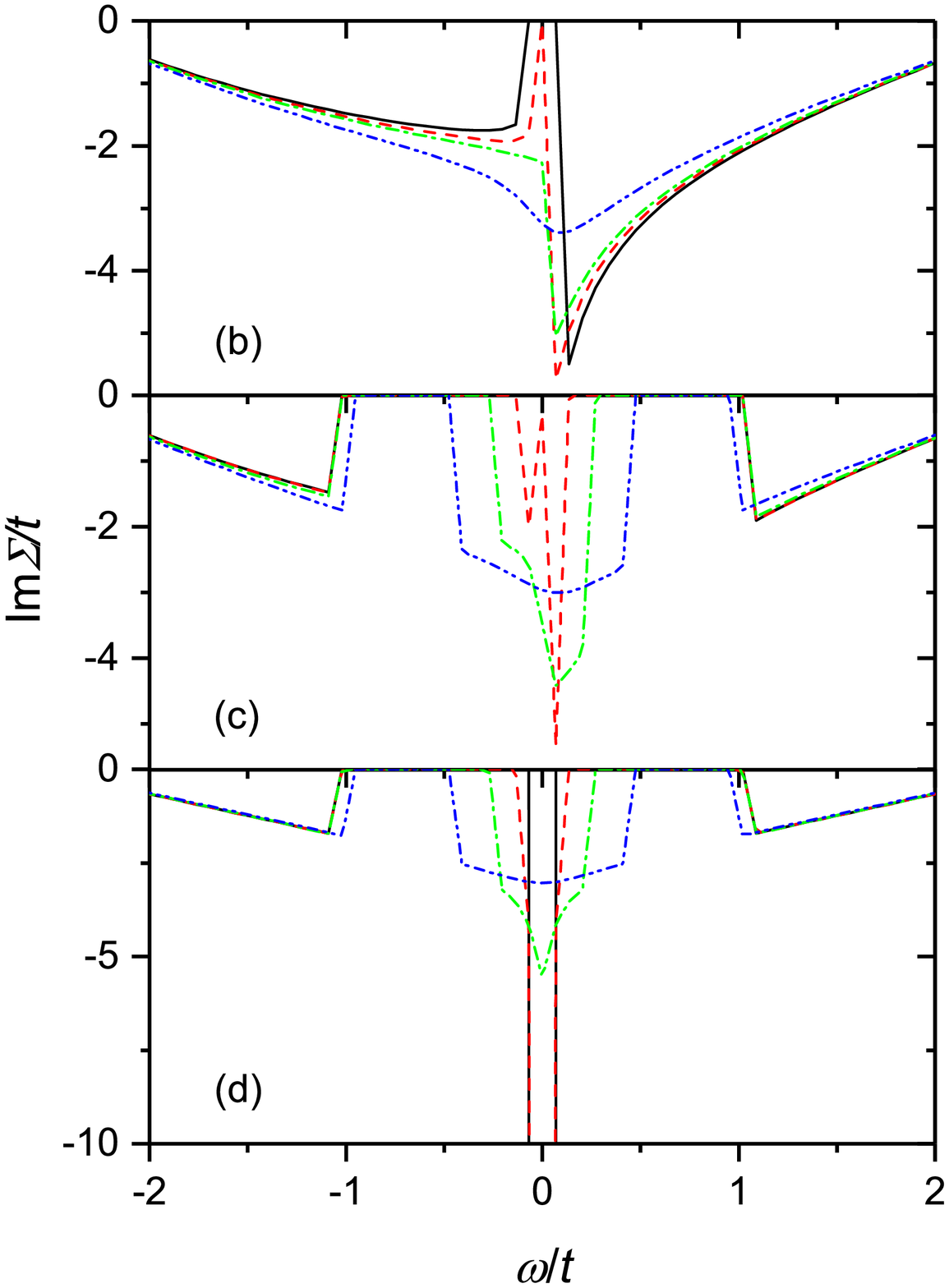}}}
\caption{The change of the imaginary part of the self-energy at crossing the separation curve in Fig.~\protect\ref{Fig7}. $U=5.1t$. (a) $\eta=0.68t$, ${\bf k}=(0,0)$ and $T=0.578t$ (black solid line), $T=0.605t$ (red dashed line), $T=0.633t$ (green dash-dotted line) and $T=0.825t$ (blue dash-dot-dotted line). (b) The part of the pane (a) near the Fermi level in a larger scale. (c) The self-energy near the Fermi level for $\eta=0.408t$, ${\bf k}=(0,0)$ and $T=0.611t$ (black solid line), $T=0.616t$ (red dashed line), $T=0.66t$ (green dash-dotted line) and $T=1.101t$ (blue dash-dot-dotted line). (d) The same as in pane (c), but for ${\bf k}=(\pi/2,\pi/2)$.} \label{Fig8}
\end{figure}
Let us consider the self-energy of the obtained solutions,
\begin{equation}\label{selfenergy}
\Sigma({\bf k}\omega)=\omega-t_{\bf k}-G^{-1}({\bf k}\omega)=\omega-K^{-1}({\bf k}\omega).
\end{equation}
The variation of its imaginary part at the crossing of the separation curve in Fig.~\ref{Fig7} is shown in Fig.~\ref{Fig8}. On the border of the magnetic Brillouin zone ${\rm Im}\Sigma$ diverges at $\omega=0$ in the insulating state due to simultaneous vanishing of real and imaginary parts of Green's function (the black solid line in Fig.~\ref{Fig8}(d), which shows only the foot of the peak). An intensive peak at these frequency and wave vectors is retained also in metallic states near the separation curve. For this and other wave vectors there is nothing resembling the Fermi liquid behavior in self-energies of metallic states. It is not surprising because for the considered moderate Hubbard repulsions the metallic states exist at comparatively high temperatures, and the decrease of $T$ leads to the transition to insulating states (see Fig.~\ref{Fig7}). As mentioned above, this our result differs from the respective DMFT outcome where there is a range of $U$, in which the temperature decrease leads to transition from insulator to metal \cite{Georges}. The Fermi liquid behavior is inherent in small Hubbard repulsions. Though the strong coupling diagram technique gives the correct result in the limit $U\rightarrow 0$ \cite{Vladimir}, its applicability for this region is questionable.

\begin{figure}[tbh]
\centerline{\resizebox{0.8\columnwidth}{!}{\includegraphics{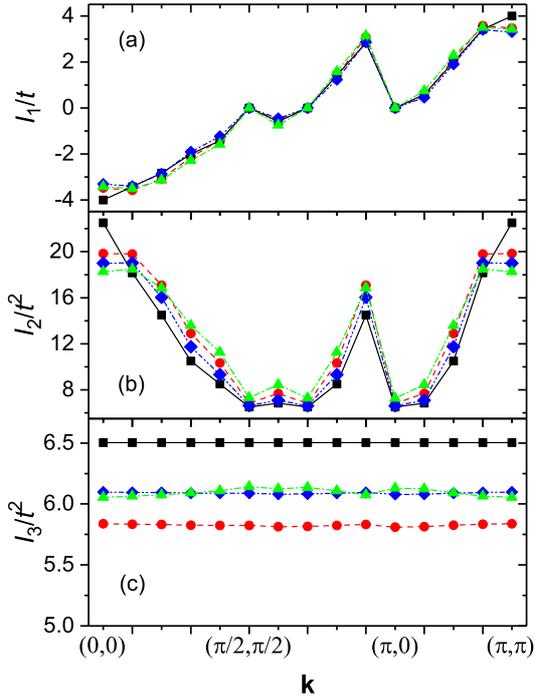}}}
\caption{Values of the integrals in the sum rules~(\protect\ref{I1})-(\protect\ref{I3}) for 15 wave vectors in the octant $(0,0)-(\pi,\pi)-(\pi,0)$ of the Brillouin zone for $U=5.1t$. Red circles and dashed curves correspond to $T=1.101t$ and $\eta=0.408t$, green triangles and dash-dotted curves are for $T=0.55t$ and the same $\eta$, blue rhombuses and dash-dot-dotted curves correspond to $T=1.101t$ and $\eta=1.1t$. Right-hand sides of Eqs.~(\protect\ref{I1})-(\protect\ref{I3}) are shown by black squares and solid lines in the respective panels of the figure.} \label{Fig9}
\end{figure}

In conclusion let us examine the fulfilment of sum rules for the above solutions. For the considered one-particle Green's function the following sum rules \cite{Kalashnikov,White,Vilk} can be inspected:
\begin{eqnarray}
{\cal I}_0&\equiv&\int_{-\infty}^{\infty}A({\bf k}\omega)d\omega=1,\label{I0}\\
{\cal I}_1&\equiv&\int_{-\infty}^{\infty}\omega A({\bf k}\omega)d\omega=t_{\bf k}-\mu+U\langle n_{\bf l,-\sigma}\rangle,\label{I1}\\
{\cal I}_2&\equiv&\int_{-\infty}^{\infty}\omega^2 A({\bf k}\omega)d\omega=(t_{\bf k}-\mu)^2+2U(t_{\bf k}-\mu)\langle n_{\bf l,-\sigma}\rangle\nonumber\\
&&\quad+U^2\langle n_{\bf l,-\sigma}\rangle,\label{I2}\\
{\cal I}_4&\equiv&-\int_{-\infty}^{\infty}{\rm Im}\Sigma({\bf k}\omega)d\omega=\pi U^2\langle n_{\bf l,-\sigma}\rangle\nonumber\\
&&\quad\times(1-\langle n_{\bf l,-\sigma}\rangle),\label{I3}
\end{eqnarray}
where $\langle n_{\bf l,-\sigma}\rangle$ is the mean value of the occupation number, which is equal to $1/2$ at half-filling. Hence in this case the above integrals should not depend on the temperature and artificial broadening $\eta$. Besides, the first and last integrals have to be independent of ${\bf k}$. Calculations show that for the obtained solutions ${\cal I}_0$, Eq.~(\ref{I0}), may have a deviation within the limits of 30\% from unity. Therefore, spectral functions were scaled to fulfil this sum rule before calculating integrals in Eqs.~(\ref{I1}) and (\ref{I2}). The values of these integrals and the right-hand sides of Eqs.~(\ref{I1}) and (\ref{I2}) are compared in Fig.~\ref{Fig9}(a,b) for all 15 momenta in the octant $(0,0)-(\pi,\pi)-(\pi,0)$ of the Brillouin zone.
As seen from the figure, the sum rules are obeyed quite satisfactorily. The integrals depend on temperature and broadening only slightly. In Fig.~\ref{Fig9}(c) the left- and right-hand sides of Eq.~(\ref{I3}) are compared for the same parameters. The integrals are nearly independent of momentum as it must, and their values are close to $U^2/4$. The temperature dependence is weak, while the dependence on $\eta$ is somewhat stronger.

\section{Conclusion}
\label{Concl}
In this work, the 2D repulsive one-band Hubbard model at half-filling was investigated using the strong coupling diagram technique. Together with the local terms the lowest-order nonlocal term was included into the irreducible part. The obtained system of equations for the electron Green's function was solved by iteration for moderate Hubbard repulsions and temperatures. As the initial Green's function in this procedure we used the result of the Hubbard-I approximation, poles of which were shifted to the lower frequency half-plane to obtain the proper analytic behavior for the retarded function. It was found that obtained solutions are changed with the variation of this shift $\eta$. The changes are mainly located in the frequency region near the Fermi level -- in insulating solutions the width of the Mott gap is affected, while in metallic solutions the shape of the spectral function is modified. Scenarios of the metal-insulator transition are also changed with $\eta$ -- for its smaller values the transition occurs due to sudden appearance of a narrow band of mobile states in the middle of the Mott gap, while for larger $\eta$ the transition happens with gradual closure of the gap as the temperature increases. Hence for given values of the Hubbard repulsion and temperature the derived equations have continua of metallic or insulating solutions. The plurality of the solutions is caused by the nonlocal term -- in the similar procedure with the irreducible part constructed from local terms solutions were independent of $\eta$. Evidently these solutions correspond to different values of the grand potential, and the continuous variation of their spectral shapes with $\eta$ implies continuous dependence of the potential on this parameter. In its turn, this suggests the lack of a gap between solutions with the smallest and larger potentials.

In spite of the multiplicity of the solutions, they have a common curve separating metallic and insulating states in the $U$-$T$ phase diagram. This curve separates solutions obtained by iteration from the Hubbard-I Green's function. One can start iteration from one of these solutions and, varying the temperature in iteration, obtain a metallic solution in the insulating region or, conversely, an insulating solution in the metallic region. Thus, not only different metallic or insulating solutions coexist at given $U$ and $T$, but also distinct types of solutions go together in the vicinity of the separating curve. This result resembles the behavior of solutions near the first-order transition line in DMFT and cluster methods, however in our case, in view of the mentioned continuity, it would be better to say about two competing continua of states rather than two states. Grand potentials of the obtained solutions were not calculated in this work and, therefore, we cannot identify the separation curve with the critical line. However, they can be supposed to be close. In our phase diagram metallic states are located at smaller Hubbard repulsions and higher temperatures, while insulating states are at higher $U$ and smaller $T$. This layout is similar to that found in some cluster approaches and in contradiction with DMFT. In accordance with our phase diagram metallic states exist at comparatively large temperatures for moderate Hubbard repulsions. Therefore, nothing resembling the Fermi liquid behavior is observed in self-energies of these solutions. Spectral functions of both metallic and insulating solutions demonstrate intensity suppressions at frequencies $\omega=\pm U/2$, which were found also in the lower-order approximation and identified with the spectral redistribution due to multiple reabsorption of carriers with the creation of states with double site occupancies. These pseudogaps lead to the distinctive four-band structure of spectral functions, which was noticed earlier in Monte Carlo simulations. Obtained solutions were found to obey quite satisfactorily sum rules for the one-particle Green's function.

In the used approximation, the obtained solutions are stationary many-particle states, into which a crystal can be brought in the course of preparation. We argued that these states transform into long-living metastable states under perturbations in view of their many-particle character and strong electron correlations. The continuity of the solutions, caused by the nonlocal term, suggests that the inclusion of unaccounted nonlocal diagrams will transform these states into a continuum of somewhat different stationary or long-living me\-ta\-sta\-ble states.

\begin{acknowledgement}
This work was supported by the research project IUT2-27 and the Estonian Scientific Foundation (grant ETF-9371).
\end{acknowledgement}

\end{document}